\documentclass[12pt,
aps,prl,
reprint,
floatfix,
]{revtex4-2}

\usepackage{upgreek}
\usepackage{bm}
\usepackage{xr}
\usepackage{graphicx}
\usepackage{gensymb}
\usepackage{xfrac}
\usepackage{extdash}
\usepackage{epsfig}
\usepackage{isomath}
\usepackage{textcomp}
\usepackage{bm}
\usepackage[english]{babel}
\usepackage{todonotes}
\usepackage{color,soul}
\usepackage{times}
\usepackage{mathtools}
\usepackage[separate-uncertainty=true]{siunitx}
\DeclareFontFamily{U}{euc}{}
\DeclareFontShape{U}{euc}{m}{n}{<-6>eurm5<6-8>eurm7<8->eurm10}{}%
\DeclareSymbolFont{AMSc}{U}{euc}{m}{n} 
\DeclareMathSymbol{\umu}{\mathord}{AMSc}{"16} 
\newcommand{\Mum}{\ensuremath{\umu}\ensuremath{\mathrm{m}}}
\newcommand{\mum}{\textrm{\,\ensuremath{\mathrm{\Mum}}}}

\mathchardef\ordinarycolon\mathcode`\:
\mathcode`\:=\string"8000
\begingroup \catcode`\:=\active
  \gdef:{\mathrel{\mathop\ordinarycolon}}
\endgroup

\begin{document}

\title{Imaging ablator-fuel mix of hot spot in inertial confinement fusion via resonant X-ray absorption using an X-ray free electron laser}
\author{Lingen Huang$^{1,8}$}
\email{lingen.huang@hzdr.de}
\author{Long Yang$^{1,8}$}
\author{Alejandro Laso Garcia$^{1}$}
\author{Oliver S. Humphries$^{2}$}
\author{Michal Šmíd$^{1}$}
\author{Mikhail Mishchenko$^{2}$}
\author{Victorien Bouffetier$^{1}$}
\author{Carsten Baehtz$^{1}$}
\author{Erik Brambrink$^{2}$}
\author{Samuele D. Di Dio Cafiso$^{1}$}
\author{Gabriel Pérez-Callejo$^{3}$}
\author{Yangzhe Cui$^{1}$}
\author{Thea Engler$^{4}$}
\author{Wen Feng$^{1}$}
\author{Sebastian G\"{o}de$^{2}$}
\author{Joerg Grenzer$^{1}$}
\author{Christian Gutt$^{5}$}
\author{Johannes Hagemann$^{4}$}
\author{Marie-Luise Herbert$^{1,6}$}
\author{Arthur Hirsch-Passicos$^{1}$}
\author{Hauke H\"{o}ppner$^{1}$}
\author{Thomas Kluge$^{1}$}
\author{Dominik Kraus$^{1,6}$}
\author{Julian Lütgert$^{6}$}
\author{Masruri Masruri$^{1}$}
\author{Motoaki Nakatsutsumi$^{2}$}
\author{\"{O}zg\"{u}l \"{O}zt\"{u}rk$^{5}$}
\author{Paweł Ordyna$^{1,7}$}
\author{Xiayun Pan$^{1}$}
\author{Franziska Paschke-Bruehl$^{1,7}$}
\author{Alexander Pelka$^{1}$}
\author{Thomas R. Preston$^{2}$}
\author{Chongbing Qu$^{6}$}
\author{S.V. Rahul$^{2}$}
\author{Lisa Randolph$^{5}$}
\author{Martin Rehwald$^{1}$}
\author{Hans-Peter Schlenvoigt$^{1}$}
\author{Samuel Schumacher$^{6}$}
\author{Ulrich Schramm$^{1,7}$}
\author{Jan-Patrick Schwinkendorf$^{1}$}
\author{Georgiy Shoulga$^{1}$}
\author{Monika Toncian$^{1}$}
\author{Jan Vorberger$^{1}$}
\author{Karl Zeil$^{1}$}
\author{Ulf Zastrau$^{2}$}
\author{Toma Toncian$^{1}$}
\author{Thomas E. Cowan$^{1,7}$}
\email{t.cowan@hzdr.de}

\affiliation{
1. Helmholtz-Zentrum Dresden-Rossendorf, 01328 Dresden, Germany\\
2. European XFEL, 22869 Schenefeld, Germany\\
3. Departamento de Física Teórica, Atómica y Óptica, Universidad de Valladolid, Valladolid 47011, Spain\\
4. Center for X-ray and Nano Science CXNS, Deutsches Elektronen-Synchrotron DESY, Notkestraße 85, 22607 Hamburg, Germany \\
5. Universit\"{a}t Siegen, 57072 Siegen, Germany\\
6. Universit\"at Rostock, 18057 Rostock, Germany\\
7. Technische Universit\"at Dresden, 01062 Dresden, Germany\\
8. These authors contributed equally to this work.\\
}

\date{\today}

\begin{abstract}

A novel diagnostic is proposed for active probing of doped inertial confinement fusion (ICF) spherical shells, enabling direct measurements of mix and burn dynamics at next-generation implosion facilities via resonant absorption imaging with an X-ray free-electron laser (XFEL). The charge-state-sensitive imaging of embedded dopants provides spatiotemporally resolved constraints on ionization, opacity, and material mix seeded by hydrodynamic instabilities in stagnated plasmas. Proof-of-principle experiments demonstrating resonant XFEL probing of hot spot in coated copper wires driven by an ultra-short relativistic laser pulse, have established the feasibility of this approach. Furthermore, atomic and radiation-hydrodynamic simulations, combined with synthetic resonant X-ray absorption imaging, extend the diagnostic concept to laser-driven direct-drive ICF shells with copper-doped ablators and radiation-driven indirect-drive shells with tungsten-doped ablators. This unique approach, combining a laser-driven implosion facility with a high-brightness XFEL, could enable precise measurements of fusion-relevant hot dense plasmas at multi-keV temperatures, provide stringent benchmarks for radiation-hydrodynamics models, and advance the realization of inertial fusion energy.

\end{abstract}

\maketitle 

\section{Introduction}

Since the successful demonstration of fusion ignition and burn in 2022 at the National Ignition Facility (NIF)~\cite{Abu2022,Abu2024}, interest in pursuing laser-driven inertial fusion energy (IFE) as a clean, low--carbon and sustainable energy source has been re-invigorated and grown significantly. One of the most critical scientific challenges on the path toward viable IFE power plant is to identify high-gain schemes that are sufficiently robust. In practice, however, the conditions required for high gain—extreme compression, strong heating, and precise symmetry—impose fundamental limitations~\cite{Dewald2020}. Hydrodynamic instabilities arise when a dense shell is accelerated by lower-density plasma, as in the ablative Rayleigh--Taylor instability, or when interface perturbations are seeded by shocks, surface roughness, fill-tube perturbations, and capsule defects, leading to Richtmyer--Meshkov and Rayleigh--Taylor growth~\cite{Pickworth2018,Weber2020}. These instabilities cause small perturbations to grow rapidly and drive material mix at the fuel–ablator interface, which degrades the hot spot temperature,  stagnation compression and thus fusion performance. These effects are compounded by implosion asymmetries due to non-uniform laser drive imprint or target imperfections~\cite{Callahan2018,Braun2022}, further reducing compression efficiency. In the direct-drive case, laser–plasma interactions such as stimulated Raman scattering (SRS) and stimulated Brillouin scattering (SBS) can scatter incident energy and preheat the fuel, further lowering achievable areal densities~\cite{Theobald2012,Depierreux2019,Lei2024}. Together, these coupled limitations—instability growth, material mix, asymmetry, and energy loss—make high-gain designs intrinsically sensitive to imperfections of the initial conditions of laser–target interactions. Moreover, probing the early-time seeding and growth of hydrodynamic instabilities in inertial confinement fusion (ICF) implosions is not accessible with the current laser-driven X-ray backlighters, such as those employed in the indirect-drive scheme at the NIF and the direct-drive scheme at the OMEGA laser facilities, due to their limited resolutions~\cite{Rygg2014,Tommasini2017,Hall2019}, thereby hampering our understanding and optimizing the target design for commercial fusion energy~\cite{Weber2026}. 

Tracer elements such as  silicon (Si), copper (Cu), germanium (Ge), bromine (Br) and tungsten (W) are typically used as dopants in inertial confinement fusion experiments to generate characteristic X-ray emission for diagnosing plasma conditions~\cite{Dittrich1999,Cohen2004,Regan2012,Wang2017, BerzakHopkins2018}. Inner-shell emission spectroscopy, including K$\alpha$ and satellite lines from dopant elements embedded within the abaltors, enables the determination of electron temperature, density, and ionization balance through collisional–radiative modeling. In addition, dopants incorporated into ICF targets can mitigate fuel preheat while enhancing diagnostic sensitivity to material mix seeded by the hydrodynamic instabilities~\cite{Bachmann2020,Bachmann2022,Bachmann2023,Hall2024,Tommasini2025}. In particular, combining crystal-backlighter imaging (CBI) with a single-line-of-sight camera at NIF has demonstrated a state-of-the-art capability to detect the mix zone of W-doped high-density-carbon (HDC) ablators, but is limited to spatial and temporal resolutions of $\sim$12~$\mum$ and $\sim$35~ps, respectively~\cite{Hall2024}. These limitations motivate the development of advanced X-ray probing techniques with higher spatiotemporal resolution for future ICF experiments. Over the last decade, the advent of ultrashort, high-brilliance X-ray free-electron lasers (XFELs), delivering $\sim10^{12}$ photons per pulse with pulse durations of tens of femtoseconds and narrow spectral bandwidths, has provided unprecedented opportunities to probe transient warm and hot dense plasmas~\cite{Kluge2018,Garcia2024, Huang2026-2,Kraus2025}. Coupling an XFEL to a laser-driven implosion facility could further extend these capabilities to extreme states of matter, enabling nanometer-scale spatial and femtosecond temporal resolution and opening new avenues for X-ray diagnostics of laser-fusion plasmas. 

In this work, we propose for active probing the ablator-fuel mix using resonant X-ray absorption imaging as a charge-state-sensitive diagnostic, enabled by an extremely bright XFEL backlighter. This approach is motivated by our recent demonstration of resonant X-ray emission spectroscopy (RXES) to track the ionization and recombination dynamics in a Cu wire irradiated by a relativistic-intensity laser pulse~\cite{Huang2026,Huang2026-2}. The measurements clearly revealed a rise-and-fall evolution of the resonant emission yield, anti-correlated with the X-ray transmission~\cite{Huang2026-2}, reflecting the temporal evolution of the heated region. Although propagation-based X-ray phase-contrast imaging (XPCI) enabled retrieval of the total X-ray transmission, reconstruction of spatially resolved phase and attenuation maps remains challenging because phase-shift and absorption effects are intrinsically coupled and imaging quality is limited in the diffractive regime~\cite{Huang2026}. To overcome these limitations, here we demonstrate spatially resolved measurements of density and attenuation in coated Cu wires driven by a relativistic laser using the newly commissioned Talbot X-ray imaging system at the European-XFEL facility~\cite{LasoGarcia2026}. A two-dimensional phase diffraction grating placed in the X-ray beam generates a self-image at a well-defined Talbot distance, from which the absorption and phase distributions of the hot dense plasma are independently reconstructed, revealing the spatiotemporal evolution of the hot spot. We further propose extending this capability to compressed fuel in ICF research, enabling the investigation of ablator-fuel mix at next-generation implosion facilities coupled to an XFEL. By tuning the incident XFEL photon energy to selected absorption transitions, the high spectral brightness enables resonant excitation of tracer elements embedded within mid-Z or high-Z-doped ablator layers of fusion shells. Aided by magnetohydrodynamic (MHD) simulations, synthetic X-ray absorption images are generated for both direct- and indirect-drive configurations using an extended two-color XFEL scheme, illustrating how complementary on- and off-resonance probing can enhance the diagnostic capabilities for inertial fusion experiments. This spectral selectivity enables localized imaging of mix dynamics within specific regions of the target, providing sensitivity to spatially and compositionally heterogeneous plasmas under extreme compression and heating conditions.

\section{Proof of principle experimental results of resonant XFEL probing of hot spot in coated Cu wires}

\begin{figure*}[htb]
    \centering
    \includegraphics[width=\textwidth]{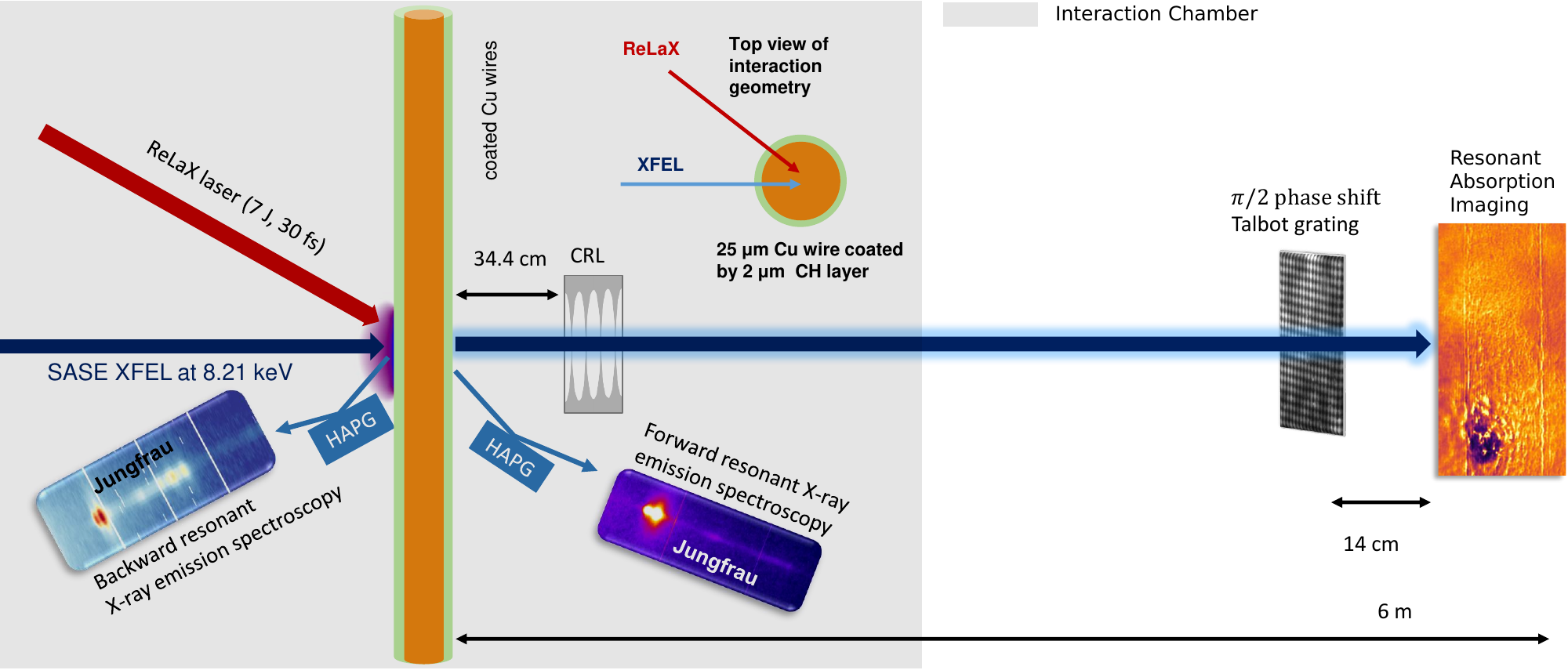}
    \caption{A schematic of the experimental setup, illustrating the optical laser pump–XFEL probe geometry and diagnostic configuration.}
    \label{fig1}
\end{figure*}

\begin{figure*}[htb]
    \centering
    \includegraphics[width=\textwidth]{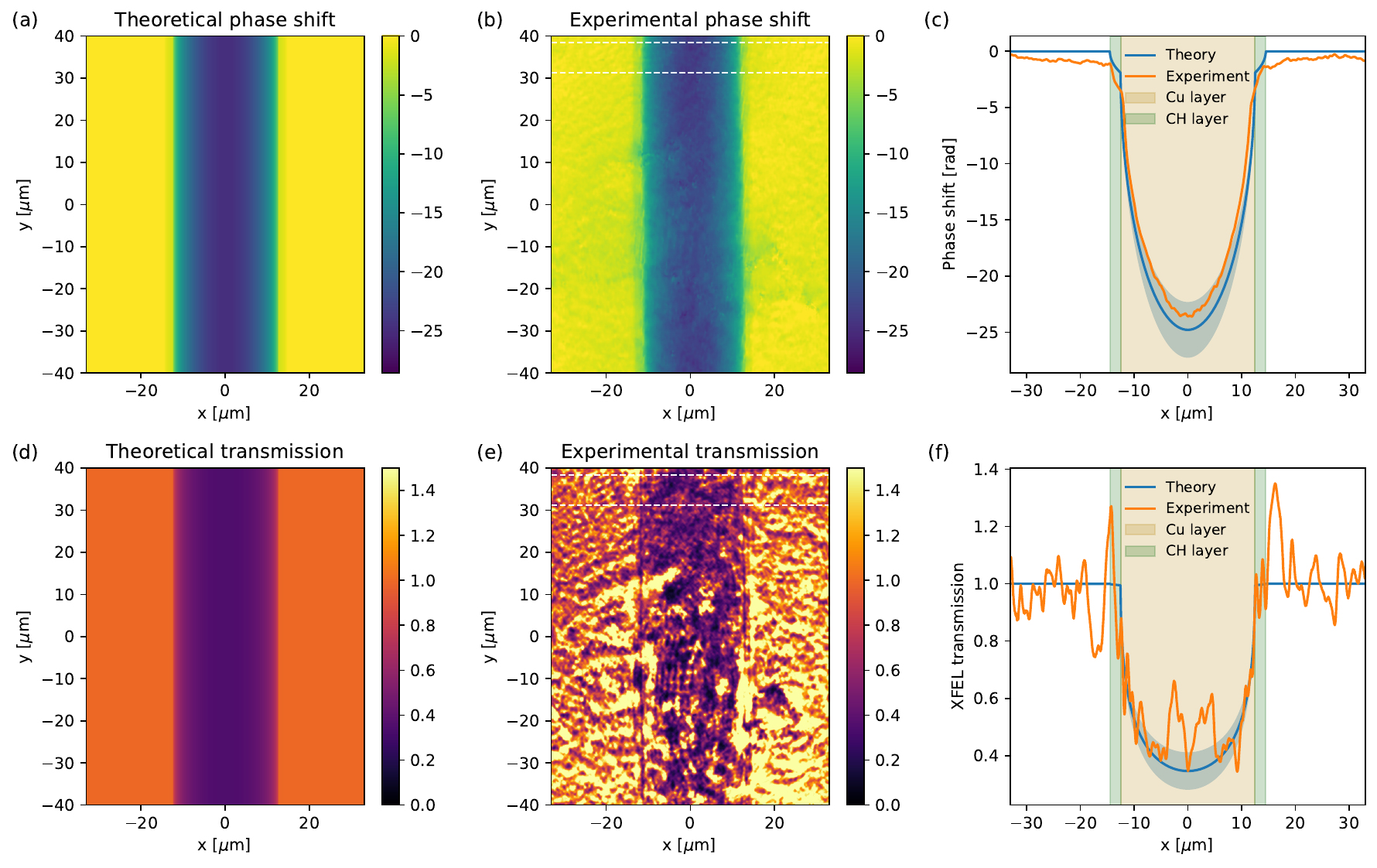}
    \caption{Reconstructed phase and transmission maps for cold coated Cu wire targets, with comparison to theory. The light-green and brown shaded regions in (c) and (f) indicate the outer CH and inner Cu layers, respectively. The shaded theoretical phase shift and transmission account for a 10\% uncertainty in the thickness of the 25~$\mum$ Cu wire.}
    \label{fig2}
\end{figure*}

\begin{figure*}
    \centering
    \includegraphics[width=\textwidth]{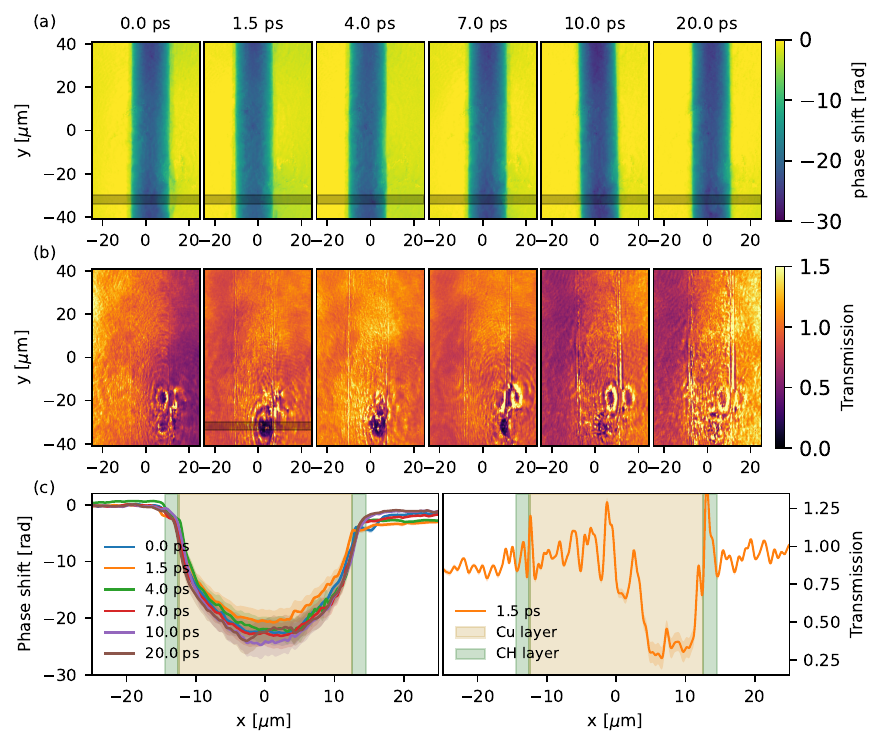}
    \caption{Time-resolved reconstructions of (a) the phase and (b) the transmission, normalized to the cold coated Cu wire target, together with (c) the corresponding lineouts around the laser–wire interaction region. Phase maps are shown for delays from 0 to 20~ps, while the transmission map is shown at a representative delay of 1.5~ps. The gray shaded areas in (a) and (b) indicate the regions over which the lineouts are averaged, while the light-green and brown shaded regions in (c) denote the nominal inner Cu and outer CH layers, respectively.} 
    \label{fig3}
\end{figure*}

The proof-of-principle experiment was conducted at the HED-HiBEF instrument at the European XFEL, where the ultra-intense ReLaX laser generated solid-density plasmas and the XFEL served as a backlighter for imaging the plasma dynamics. A schematic of the experimental setup is shown in Fig.~\ref{fig1}, illustrating the optical laser pump–XFEL probe geometry and diagnostic configuration. The ReLaX laser is a Ti:sapphire-based system delivering pulses of up to 7 J with a 30 fs full width at half maximum (FWHM) duration, focused to a 4~$\mum$ (FWHM) spot on a 25 $\mum$ Cu wire coated with a 2 $\mum$ CH layer. The XFEL probe beam, generated via self-amplified spontaneous emission (SASE), delivered 1.7~ mJ in 25 fs (FWHM) pulses with a 20~ eV (FWHM) bandwidth. The photon energy was tuned to 8.21 keV, in resonance with the K–L transition in nitrogen-like Cu$^{22+}$ ions in hot solid density plasmas. The XFEL beam illuminates the wire target over a $250\mum \times 250 \mum$ region. The object plane was imaged and magnified using a compound refractive lens (CRL) stack of 15 beryllium lenses (f $\approx$ 34.4 cm)  onto an Andor Sona detector with a 10× objective, located 6 m downstream. This configuration yields a total magnification of 
$\mathrm{M} \approx 160$. Accordingly, the effective pixel size at the object plane is 
$6.5~\mum/\mathrm{M} \approx 40.6~\mathrm{nm}$. The resulting field of view (FoV) for the object on the detector ($2160 \times 2560$ pixels) is $85.6~\mum \times 104~\mum$. In addition, a checkerboard diamond grating with a diagonal pitch of 6.5~$\mum$ and a thickness of 10~$\mum$ was placed 14 cm in front of the detector, corresponding to a \(\pi/2\) Talbot distance for the Talbot imaging setup. Complementary backward and forward X-ray emission spectroscopy diagnostics with Highly Annealed Pyrolitic Graphite (HAPG) crystals were employed to monitor the time-integrated resonant and self-emission, and characterize the initial laser–plasma interaction conditions. 

For each wire target, a sequence of three runs was performed: a cold shot, a hot shot, and a post-shot, corresponding respectively to the XFEL-only probe on the unperturbed target, the combined ReLaX+XFEL interaction, and the XFEL probe following complete target destruction by the ReLaX pulse. In general, the X-ray propagation through the target material is governed by the complex-valued refractive index $n = 1-\delta+i\beta$, where the real and imaginary parts of the refractive index determines the phase shift of the X-ray wave and its attenuation due to absorption respectively~\cite{AlsNielsen2011}. In Talbot imaging, the X-ray intensity recorded at the detector after propagating through the target and vacuum can be expressed as~\cite{Bouffetier2020,PrezCallejo2023}:
\begin{align}
I_{\text{ref}}(r) &= A_{\text{ref}}(r) + B_{\text{ref}}(r)\, e^{i\,[k \cdot r + \Phi_{\text{ref}}(r)]} \\
I_{\text{obj}}(r) &= A_{\text{obj}}(r) + B_{\text{obj}}(r)\, e^{i\,[k \cdot r + \Phi_{\text{obj}}(r)]}
\end{align}
where the subscripts “ref” and “obj” denote the reference (cold or post-shot) and object (hot shot), respectively, and \(A(r)\), \(B(r)\), and \(\Phi(r)\) are real functions, from which the transmission and phase shift can be respectively extracted via Fourier analysis in the Talbot imaging algorithm, as shown following:  

\begin{align}
\text{Transmission} &= \frac{A_{\text{obj}}}{A_{\text{ref}}} \\
\text{Phase shift} &= \Phi_{\text{obj}} - \Phi_{\text{ref}}
\end{align}

Because the phase shift is proportional to the projected electron density~\cite{Hagemann2021,LasoGarcia2026}, while the transmission is sensitive to the temperature-dependent opacity~\cite{Huang2026-2}, the density and temperature distributions can be reconstructed simultaneously with submicrometer spatial resolution using the Talbot imaging diagnostic. This enables direct visualization of the spatiotemporal evolution of the laser-driven hot spot.

Based on the CXRO atomic database~\cite{Henke1993}, the refractive-index parameters of cold solid CH ($\rho_{\mathrm{CH}}=0.94~\mathrm{g/cm^{3}}$) and Cu ($\rho_{\mathrm{Cu}}=8.96~\mathrm{g/cm^{3}}$) are $\delta_{\mathrm{CH}}=3.12142\times10^{-6}$, $\beta_{\mathrm{CH}}=4.247\times10^{-9}$, $\delta_{\mathrm{Cu}}=2.3321\times10^{-5}$, and $\beta_{\mathrm{Cu}}=5.0870\times10^{-7}$, at X-ray photon energy of 8.2 keV respectively. The phase shift and transmission accumulated during propagation through the coated wire with a CH layer of thickness $d_{\mathrm{CH}}$ and a Cu layer of thickness $d_{\mathrm{Cu}}$ is then given by
$\Delta\phi =
-k_0\left(
\delta_{\mathrm{Cu}}d_{\mathrm{Cu}}
+\delta_{\mathrm{CH}}d_{\mathrm{CH}}
\right)$, and $T =
\exp\left[
-2k_0\left(
\beta_{\mathrm{Cu}}d_{\mathrm{Cu}}
+\beta_{\mathrm{CH}}d_{\mathrm{CH}}
\right)
\right]$ respectively, where $k_0$ is the X-ray wavenumber. 
It is noteworthy that the imaginary components of the refractive index are substantially smaller than the corresponding real components for a cold target, with $\beta_{\mathrm{Cu}}/\delta_{\mathrm{Cu}}\approx 2.2\times10^{-2}$ and $\beta_{\mathrm{CH}}/\delta_{\mathrm{CH}}\approx 1.4\times10^{-3}$. Consequently, the phase shift provides substantially greater sensitivity to the material distribution than absorption at the X-ray energy considered here, making phase contrast particularly advantageous for imaging the target structure. This is evident in Fig.~\ref{fig2}, where the reconstructed phase maps exhibit excellent image quality with a high signal-to-noise ratio, whereas the transmission maps show significant fluctuations. These fluctuations are primarily attributed to phase-errors arising from fabrication imperfections and debris in the CRL stack placed between the wire and the detector. As the fluctuations are static, they could be mitigated by normalizing the hot-shot data to the cold reference shot, assuming negligible probe-beam spatial jitter. Furthermore, the refractive-index of CH is significantly smaller than that of Cu, the contribution of CH to both the phase shift and X-ray transmission is substantially weaker than that of Cu for comparable propagation distances. This pronounced difference in the refractive response enables the Cu structure to be distinguished from the surrounding CH material in phase-sensitive X-ray imaging. These features are clearly observed in the reconstructed phase and transmission maps of the cold CH-coated Cu wire target, as shown in Fig.~\ref{fig2}, which show excellent agreement with the theoretical predictions. The X-ray phase-shift and transmission maps presented in Figs.~\ref{fig2} and \ref{fig3} were reconstructed based on the TIA/TNT code~\cite{PrezCallejo2022}.

As discussed above, the imaginary part of the refractive index, $\beta$, is generally small for a cold target. In contrast, in laser-generated warm and hot dense plasmas, atomic excitation and ionization modify the electronic-shell configuration and electron screening of the nuclear Coulomb potential. For a given transition, its transition energy depends on the number of bound electrons and thus on the ionic charge state and plasma temperature. As the temperature increases, the average charge state rises, reducing electron screening and shifting the transition energy to higher photon energies. This dependence enables time-resolved probing of hot-spot evolution using a charge-state-sensitive diagnostic, for example, by tuning an XFEL into resonance with a bound–bound transition of a selected ionic species. Under resonant conditions, $\beta$ can increase substantially because the probability of photon absorption is strongly enhanced when the XFEL photon energy matches an allowed electronic transition~\cite{Huang2026-2}. Since $\beta$ is directly related to the X-ray absorption coefficient, this resonant enhancement results in substantially increased attenuation, providing element- and charge-state-sensitive contrast in absorption-based X-ray imaging. This effect is clearly demonstrated in Fig.~\ref{fig3} (a) and (b), which show the temporal evolution of reconstructed two-dimensional phase and transmission maps, normalized to the cold target, at delays ranging from 0 to 20~ps after ReLaX laser irradiation of the coated Cu wire. Over this timescale, the phase shift remains largely unchanged compared to the cold case, indicating negligible target deformation due to hydrodynamic expansion or compression, consistent with previous XPCI measurements~\cite{Garcia2024,Huang2026}. In contrast, a clear evolution of the absorption is observed at the laser-wire interaction point. A reduction in transmission by approximately a factor of four relative to the cold regions of the wire is observed at 1.5 ps with a heating and ionization spot of $\sim$7~\mum, as shown in Fig.~\ref{fig3} (c), indicating an increased population of highly ionized Cu$^{22+}$ ions, whose K-to-L transition energy is resonant with the XFEL photon energy. The highly charged ions subsequently diffuse into a larger region, with the enhanced absorption width localized around the laser focal spot before diminishing, indicating high opacity and thus ionization in the bulk target with strong spatial gradient. After 10 ps, the transmission becomes comparable to that of the cold wire, consistent with subsequen plasma cooling and recombination. It is noted that the bright, ring-like diffraction patterns surrounding the strongly attenuating region may arise from the Airy-like spatial profile of the ReLaX laser focus, imperfections in the CRL X-ray optics, or diffraction induced by the opacity gradient. Since the change in absorption (or plasma opacity) at the resonant photon energy for the selected charge state is directly correlated with temperature, the observed spatiotemporal evolution reflects the underlying transient heating process of solid density plasmas. According to atomic calculations shown in Fig.~\ref{Cu_OD}, the opacity at 8.21 keV remains nearly identical to that of cold Cu up to $\sim$200 eV, then increases to a maximum around $\sim$500 eV, and subsequently decreases with temperature due to over-ionization. On picosecond timescales, recombination during plasma cooling further reduces the opacity. The proof-of-principle resonant absorption imaging data thus provide access to the spatiotemporal evolution of key plasma parameters with unprecedented precision, including opacity, ionization, and temperature. We emphasize that the tamped Cu wire targets studied here differ from the pure Cu wires investigated in our previous experiments~\cite{Garcia2024,Yang2025,Huang2026,Huang2026-2}. In the present targets, the surrounding ablator CH tamping layer substantially suppresses the formation of an extended coronal Cu plasma, thereby enabling the isochoric heating of the buried Cu plasma. This capability can be further extended to investigate instability growth and material mix by incorporating imaging tracer elements into complex, multilayer, fusion-relevant shell targets using this novel diagnostic, as discussed in the following sections.

\section{Syntheic resonant X-ray imaging of ablator--fuel mix in ICF}

Motivated by the proof-of-principle experimental data, resonant X-ray absorption imaging using a bright, narrow-band XFEL is proposed to diagnose ablator–fuel mix seeded by hydrodynamic instabilities in doped spherical ICF shell targets. For laser direct-drive ICF, the XFEL photon energy is tuned to Cu K–L resonance transitions (8–9 keV), while tungsten L–M resonances (8–12 keV) are targeted for indirect-drive ICF (radiation-driven). For both cases, atomic collisional–radiative simulations using FLYCHK~\cite{FLYCHK2005} or SCFLY~\cite{SCFLY2007} are performed to obtain charge-state distributions and temperature-dependent opacities. Magnetohydrodynamic simulations using FLASH code~\cite{fryxell2000flash,dubey2009extensible}  are then carried out to model the compression and heating of the shells. Finally, the opacities calculated from the atomic model are coupled with the density profiles predicted by the MHD simulations to generate synthetic opacities and images for the doped-material mixing during the implosion and stagnation stage.

\subsection{Imaging Cu-doped shell in the direct-drive case}

\begin{figure*}
  \centering
  \includegraphics[width=1\textwidth]{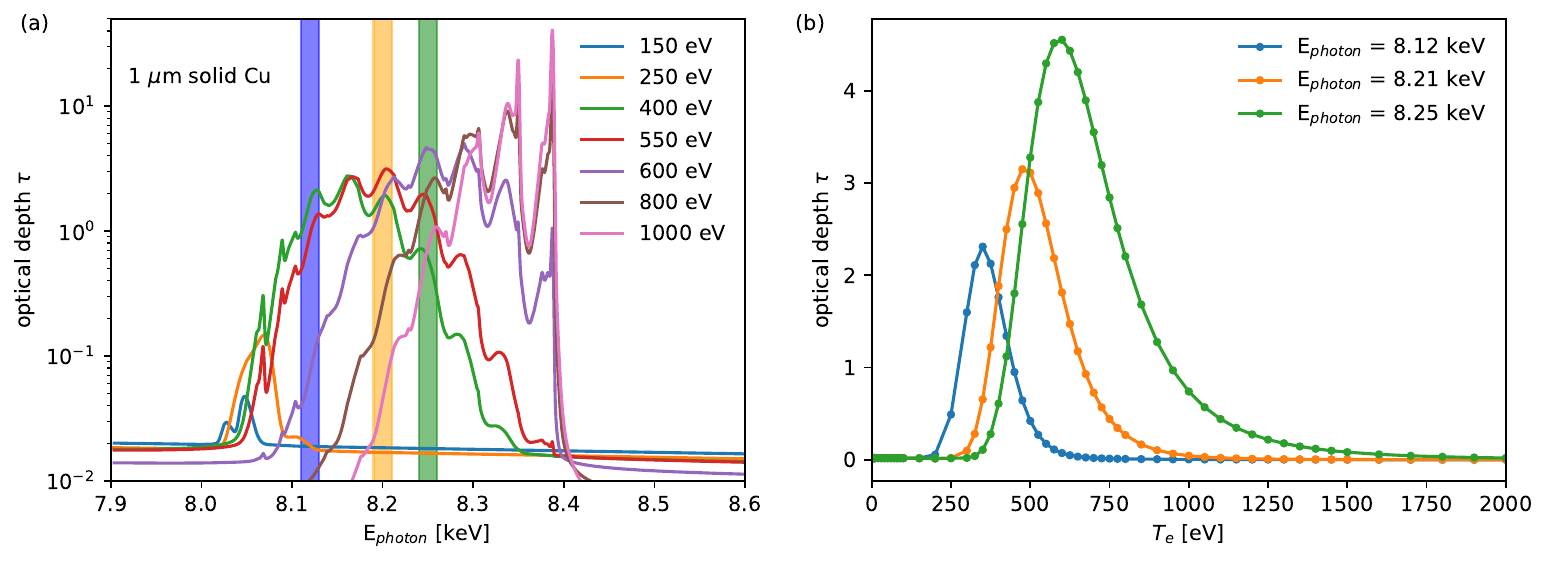}
  \caption{(a) Optical depth of a solid-density Cu plasma with 1 $\mum$ thickness at different temperatures calculated by the atomic code SCFLY over a broad photon energy range from 7.9 keV to 8.6 keV covering bound-bound and bound-free transitions for the hot dense plasma. Each peak corresponds to a specific \textit{K}-\textit{L} bound-bound transition of the Cu ions. (b) Temperature dependence of Cu's optical depth at the resonant photon energy $E_{photon}$ of 8.12 keV, 8.21 keV and 8.25 keV respectively.}
  \label{Cu_OD}
\end{figure*}

First, the optical depth $\tau$ of a solid-density Cu plasma with a thickness of $1~\mum$ is calculated over temperatures from 150 eV to 1 keV using the atomic code SCFLY~\cite{SCFLY2007}, as shown in Fig.~\ref{Cu_OD}(a). From these results, the mass attenuation coefficient $\mu$ is obtained for arbitrary density $\rho$ and path length $L$ via $\mu = \tau/(\rho L)$ and subsequently used to generate synthetic absorption images based on the MHD simulations. The opacity spectrum exhibits a set of satellite lines with partially overlapping peaks, each arising from a bound–bound transition whose energy is governed by the L-shell occupancy and corresponding Cu charge state. As the plasma temperature increases, the average ion charge state rises, leading to a shift of spectral peak positions toward higher photon energies due to reduced electron screening of the nuclear Coulomb potential. For example, X-ray photons with energies of 8.12 keV, 8.21 keV, and 8.25 keV predominantly arise from K–L bound–bound transitions in highly charged Cu$^{20+}$, Cu$^{22+}$, and Cu$^{23+}$ ions, respectively, in the solid-density Cu plasma. The corresponding temperature dependence of opacity at these selected photon energies is shown in Fig.~\ref{Cu_OD}(b). This sensitivity motivates the extension of resonant absorption imaging to ICF studies, offering a direct diagnostic of ablator–fuel mix physics.

\begin{figure*}[htb]
  \centering
  \includegraphics[width=1\textwidth]{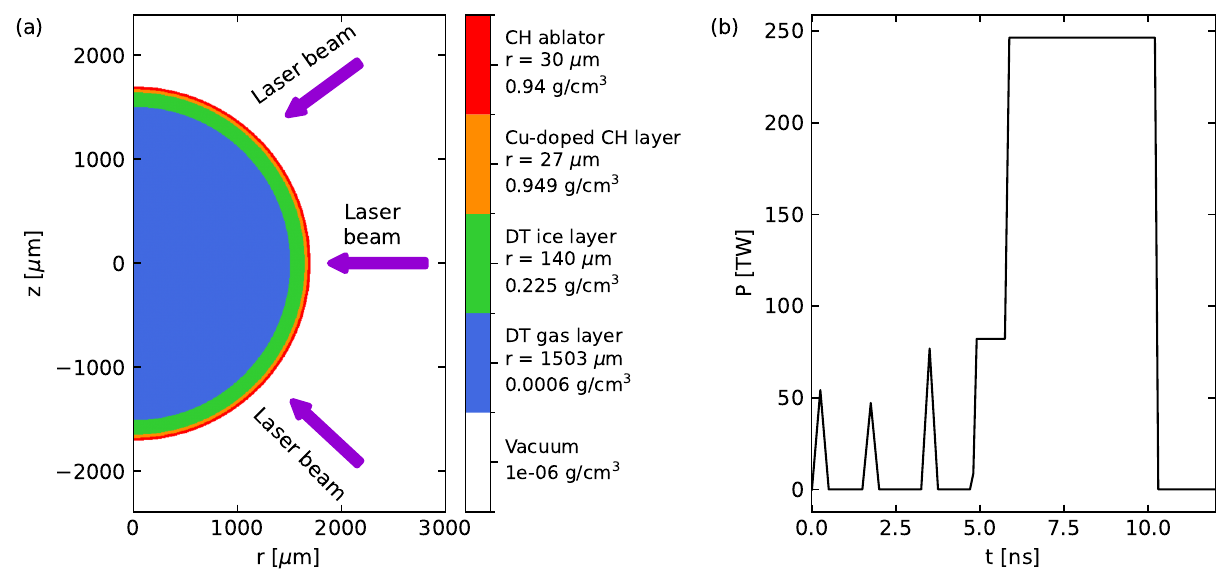}
  \caption{(a) Initial layered target geometry and laser illumination with multiple beams. (b) Temporal contrast profile of direct drive laser with triple-picket design.}
  \label{DD_Setup}
\end{figure*}

\begin{figure*}[htb]
  \centering
  \includegraphics[width=1\textwidth]{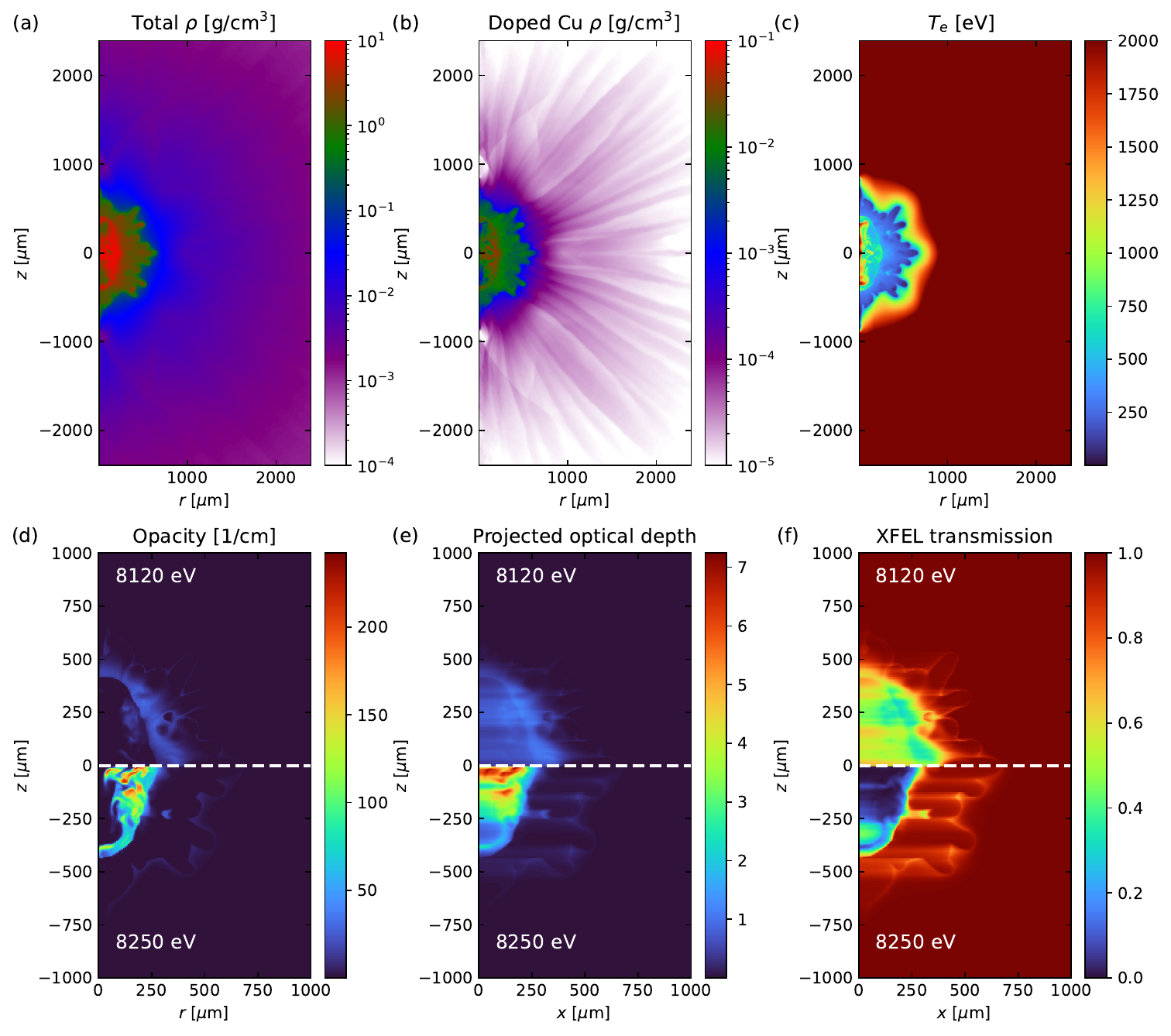}
  \caption{(a–c) Distributions of the total mass density, Cu dopant density, and plasma temperature near stagnation at 10.6 ns. (d–f) Corresponding opacity, projected optical depth, and simulated two-color X-ray transmission images at the same time for XFEL photon energies of 8.12 keV (upper panel) and 8.25 keV (lower panel) respectively.}
  \label{Cu_FLASH}
\end{figure*}

To demonstrate the diagnostic capability, an MHD simulation of direct-drive compression and implosion is performed using NIF-like laser parameters with a 1.5-MJ triple-picket design adopted from Ref.~\cite{Craxton2015}. The MHD simulations are performed using the FLASH code~\cite{fryxell2000flash,dubey2009extensible} in a two-dimensional axisymmetric $(r$–$z)$ cylindrical geometry, with the equation-of-state (EOS) data obtained from the IONMIX EOS library~\cite{Macfarlane1989}. Reflective boundary conditions are imposed along the symmetry axis, while outflow conditions are applied at the outer radial boundary. Adaptive mesh refinement is employed to resolve the steep gradients that develop during the implosion. The simulation is evolved using the one-fluid, three-temperature hydrodynamic model implemented in FLASH, with separate radiation, ion and electron temperatures.

The initial target configuration and temporal laser-power profile are shown in Fig.~\ref{DD_Setup}(a) and (b), respectively. The temporal laser-power profile consists of three initial pickets followed by the main drive pulse, which reaches a peak power of approximately $245~\mathrm{TW}$, corresponding to a peak intensity of $8\times10^{14}~\mathrm{W/cm^2}$ at the initial target surface. The layered spherical target has an initial outer radius of $1700~\mum$ and is irradiated by the laser with a wavelength of $0.35~\mum$. As shown in Fig.~\ref{DD_Setup}(a), the target consists of a $27~\mum$-thick CH ablator with a density of $0.94~\mathrm{g/cm^3}$ as the outermost layer, followed by a Cu-doped CH layer with a thickness of $30~\mum$ and a density of $0.949~\mathrm{g/cm^3}$, where 1\% Cu is added as an resonant imaging tracer dopant. These layers surround a $140~\mum$-thick cryogenic DT-ice layer with a density of $0.225~\mathrm{g/cm^3}$ and a low-density DT-gas region with a density of $0.6~\mathrm{mg/cm^3}$. The laser beams are incident on the spherical target from multiple directions, representing a spherically symmetric irradiation geometry. The laser and spherical-target parameters used in the simulations are based on the design reported in Ref.~\cite{Craxton2015}.

Figure~\ref{Cu_FLASH}(a)–(c) show the total mass density, Cu dopant density, and plasma temperature near stagnation at 10.6~ns. The simulations reveal strong modulation of the ablator–fuel interface, leading to substantial mixing between the Cu-doped CH ablator and the DT fuel, accompanied by well-developed Rayleigh–Taylor instabilities. The predicted electron temperatures, ranging from several hundred eV to keV near the core, produce highly ionized Cu charge states with L-shell vacancies, thereby enabling resonant X-ray absorption in an optical-laser pump–XFEL probe configuration. By tuning the XFEL photon energy to 8.12 or 8.25~keV, close to the relevant Cu absorption resonance, the upper and lower halves of Fig.~\ref{Cu_FLASH}(d)–(f) show the corresponding two-color opacity $\mu=\tau/L$, projected optical depth $\tau$, and simulated X-ray transmission, respectively, at 10.6~ns. The projected opacity is obtained from the simulated optical-depth distribution using an Abel transformation, assuming cylindrical symmetry about the $z$ axis. This transformation converts the radial optical-depth distribution into the corresponding line-of-sight projection, thereby providing a synthetic X-ray absorption imaging consistent with the axisymmetric geometry of the simulation. The resulting image contrast is sufficient to resolve the hydrodynamic instabilities and the associated material mixing. Importantly, the two-color probe clearly demonstrates the energy dependence of the resonant absorption response. The pronounced difference in optical depth, projected opacity, and transmission between the two photon energies highlights the enhanced absorption at the resonant energy and demonstrates the capability of two-color XFEL probing to distinguish resonant absorption from the non-resonant background. Furthermore, the ultrashort XFEL pulse duration effectively freezes the plasma evolution during the measurement, thereby minimizing motion blur. The energy-selective response therefore provides a direct signature of the resonant interaction and enables the spatially resolved identification of the mixed, highly ionized Cu-containing plasma. It should be noted that the present FLASH simulations do not include nuclear burn or $\alpha$-particle energy deposition. Consequently, the mixed Cu-doped region does not reach the multi-keV hot spot temperatures characteristic of ignition-class implosions. The results shown in Fig.~\ref{Cu_FLASH} should therefore be regarded as a design study of resonant absorption imaging in a moderately heated, hydrodynamically evolving configuration, while indicating the potential for substantially enhanced diagnostic sensitivity under ignition-relevant conditions.

\subsection{Imaging W-doped HDC shell in the indirect-drive case}

For laser indirect-drive ICF , the NIF achieved the first controlled fusion ignition to date using W-doped HDC shells~\cite{Abu2022,Abu2024}. Here, resonant X-ray absorption imaging is extended from Cu-doped direct-drive targets to W-doped shells used in indirect-drive configurations, providing a basis for feasibility studies of this diagnostic approach. Tungsten is well suited to this approach because its L-shell resonances fall within the $8$–$12$~keV photon-energy range accessible at XFEL facilities and provide strong sensitivity to the W charge-state distribution. Coupled MHD and atomic-physics simulations were used to model XFEL absorption imaging of a W-doped HDC shell in the indirect-drive configuration using two photon energies. In this scheme, resonant absorption corresponds to excitation of L-shell electrons into M-shell states. For a resonant XFEL probe at $E_{\mathrm{photon}}=8460$~eV, close to the neutral-W $L\alpha$ energy~\cite{thompson2001x}, the absorption is maximized at an electron temperature of $T_e\sim0.7$~keV, making this energy sensitive to heating and W mixed into moderately heated or preheated fuel. A second resonant channel at $E_{\mathrm{photon}}=8700$~eV provides complementary sensitivity at higher temperatures of $\sim1$~keV. The distinct temperature responses of these two channels enable the W charge-state distribution and, consequently, the thermal state of the mixed material to be probed, extending the two-color resonant absorption imaging demonstrated for the Cu-doped direct-drive case to indirect-drive ICF targets.

\begin{figure}[htb]
  \centering
  \includegraphics[width=1\columnwidth]{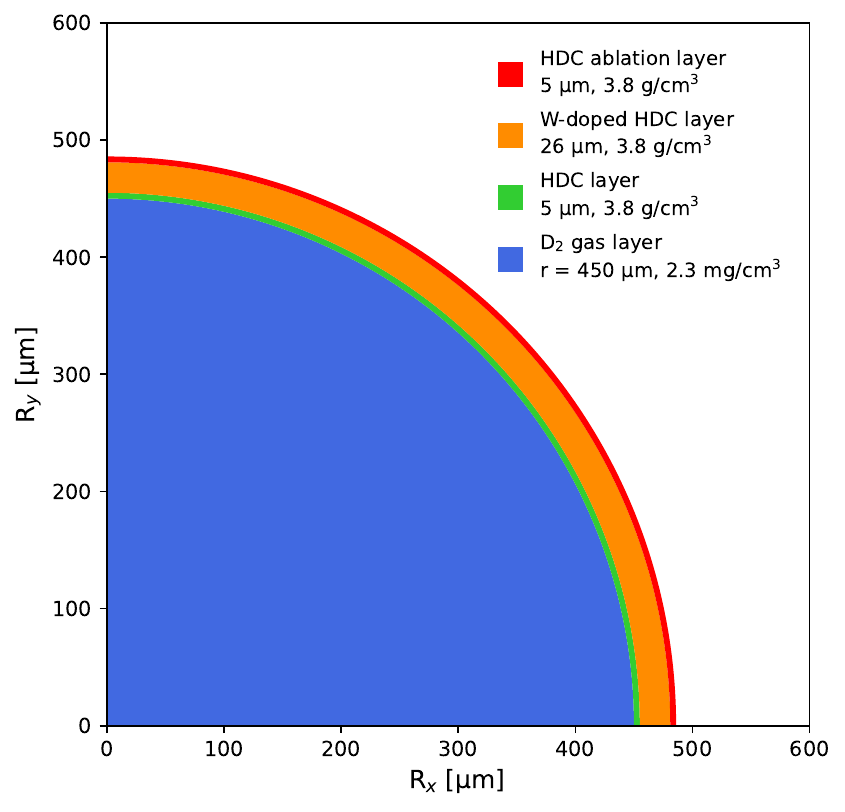}
  \caption{Initial density map of the W-doped HDC shell in the indirect-drive cas.}
  \label{all-2}
\end{figure}

To demonstrate the diagnostic feasibility, we further perform MHD simulations of an indirect-drive ICF target with a W-doped shell using a two-dimensional, quarter-symmetry cylindrical geometry. As shown in Fig.~\ref{all-2}, the initial capsule consists of a HDC ablator with an embedded W-doped HDC layer surrounding a cryogenic D$_2$ gas fill, with an initial radius of \SI{486}{\micro m}. The ablator consists of an outer \SI{5}{\micro m} layer of pure HDC, a \SI{26}{\micro m} W-doped HDC layer with a W-to-C atomic ratio of \SI{0.4}{\percent}, and an inner \SI{5}{\micro m} layer of pure HDC, enclosing a \SI{450}{\micro m} D$_2$ gas fill. The material densities are \SI{3.18}{g/cm^3} for HDC and \SI{2.3}{mg/cm^3} for D$_2$. To seed ablator--fuel interfacial mixing, an azimuthal cosine perturbation with an amplitude of \SI{2}{\micro m} is imposed on the outer HDC surface. The capsule is driven by an indirect-drive pulse with a duration of \SI{2}{ns} and a constant radiation temperature of \SI{200}{eV}. The computational domain extends to a radius of \SI{600}{\micro m}, with the capsule centered at the origin and the surrounding region initialized as vacuum. For numerical stability, the vacuum region is filled with low-density D$_2$ gas at \SI{1}{eV} temperature. The material properties are obtained from the SESAME equation-of-state tables~\cite{johnson1994sesame}. 

\begin{figure*}[htb]
  \centering
  \includegraphics[width=1.0\textwidth]{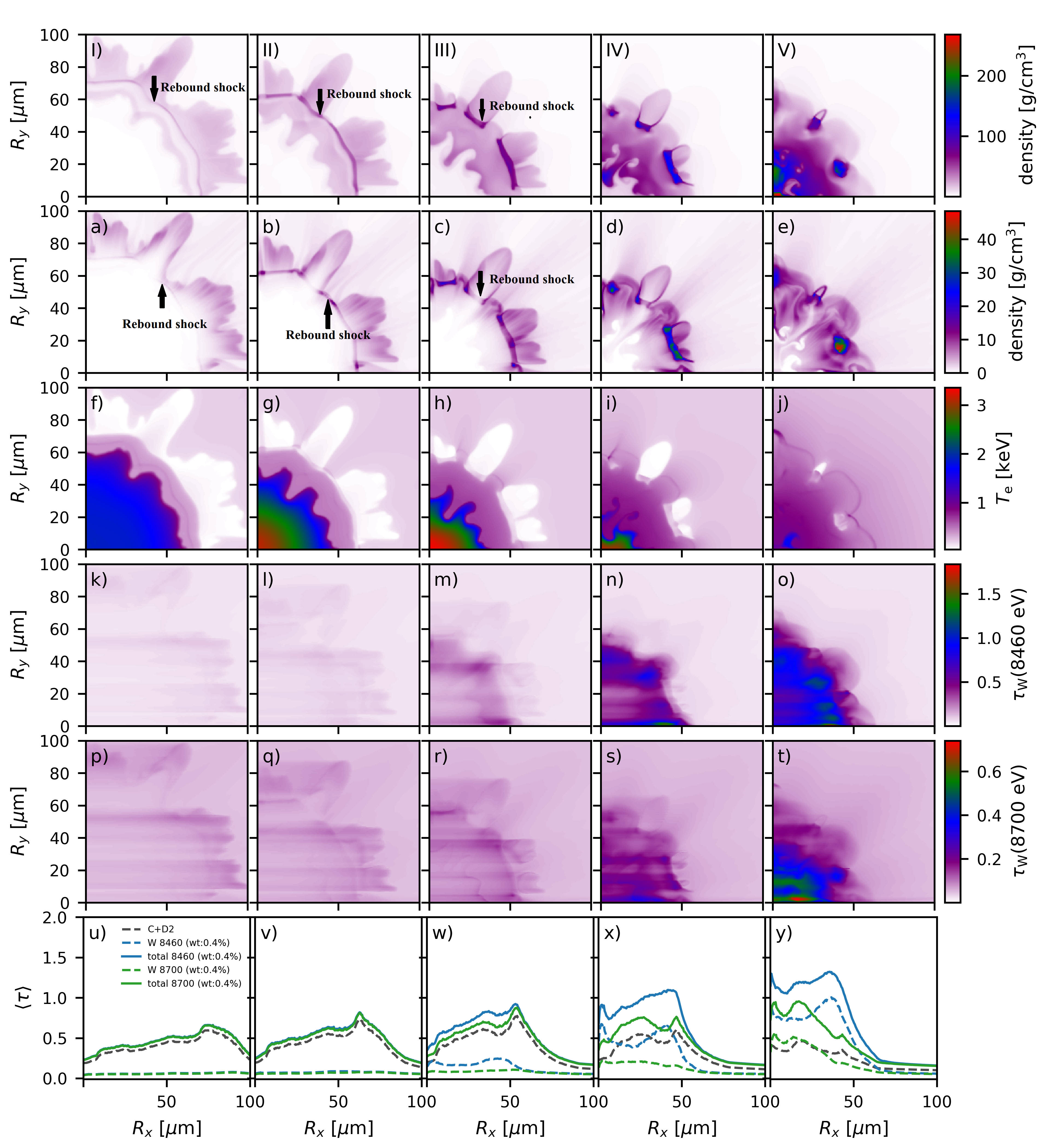}
  \caption{Time evolution of the target in the first quadrant \((R_x \ge 0,\ R_y \ge 0)\) at five selected times (from \SI{2.05}{ns} to \SI{2.25}{ns} with \SI{50}{ps} intervals) for
  (I)–(V) mass density of the total target, \(\rho\), in units of g/cm\(^3\). 
  (a)–(e) mass density of the doped layer, \(\rho\), in units of g/cm\(^3\). 
  (f)–(j) electron temperature \(T_\mathrm{e}\) in keV. 
  (k)–(o) line-of-sight tungsten optical depth \(\tau_\mathrm{W}\) at XFEL
  photon energy \(E_\gamma = \SI{8460}{eV}\). 
  (p)–(t) line-of-sight tungsten optical depth \(\tau_\mathrm{W}\) at
  \(E_\gamma = \SI{8700}{eV}\). 
  All optical depths are obtained from the 2D \((r,z)\) data assuming axial
  symmetry and integrating along the line of sight; only the quarter-plane up
  to \(R_x,R_y = \SI{100}{\micro m}\) is shown. 
  (u)–(y) corresponding azimuthlly averaged optical depth
  \(\langle \tau(R)\rangle\) in the first quadrant, plotted as a function of radius \(R\). 
  The contributions from carbon plus deuterium fuel are labeled ``C+D$_2$'',
  while the tungsten contribution is shown for the atomic fraction of 0.4\%, at both XFEL photon energies of
  \SI{8460}{eV} and \SI{8700}{eV}.}
\label{all-3}
\end{figure*}

Figures~\ref{all-3}(I)--(V) and (a)--(e) show the evolution of the total mass density and W-doped-layer density \(\rho(\mathbf{r})\), respectively, at \(t=\SI{2.05}{ns}\)--\SI{2.25}{ns} in \SI{50}{ps} increments. A rebound shock generated in the D$_2$ fuel is clearly visible as it propagates back into the W-doped HDC layer. By \(t=\SI{2.15}{ns}\), the inner HDC/D$_2$ interface becomes strongly perturbed, and the W-doped HDC begins to mix with both the inner pure-HDC layer and the D$_2$ fuel. This mixing is accompanied by pronounced Rayleigh--Taylor structures at the ablator--fuel interface, similar to those observed in the direct-drive Cu-doped shell case. The corresponding electron-temperature maps \(T_e(\mathbf{r})\) in Figs.~\ref{all-3}(f)--(j) show fast heating of the W-doped region by the rebound shock from adjacent hot fuel. The peak electron temperature in the doped layer increases from \(\sim\SI{0.5}{keV}\) at \SI{2.05}{ns} to \(\sim\SI{0.6}{keV}\) at \SI{2.15}{ns}, \(\sim\SI{0.7}{keV}\) at \SI{2.20}{ns}, and \(\sim\SI{0.8}{keV}\)--\SI{0.9}{keV} by \SI{2.25}{ns}. 

Using the simulated \(\rho(\mathbf{r})\) and \(T_e(\mathbf{r})\), we calculate the tungsten contribution to the optical depth at two XFEL photon energies, \(E_{photon}=\SI{8460}{eV}\) and \SI{8700}{eV}. Figures~\ref{all-3}(k)--(o) show the temproal evolution of tungsten optical depth \(\tau_\mathrm{W}\) at \(E_{photon}=\SI{8460}{eV}\), while Figs.~\ref{all-3}(p)--(t) show the corresponding optical depth at \(E_{photon}=\SI{8700}{eV}\). The corresponding azimuthally averaged optical depth, \(\langle\tau(R)\rangle\), over the first quadrant (\(0^\circ\)--\(90^\circ\)), is plotted as a function of radius \(R\) in Figs.~\ref{all-3}(u)--(y). Before significant C/D$_2$ mixing occurs, at \(t=\SI{2.05}{ns}\), the line-of-sight optical depth from carbon and D$_2$, \(\tau_{\mathrm{C+D_2}}\), is \(\sim 0.5\) at both probe energies, whereas the tungsten contribution is only \(\tau_\mathrm{W}\sim 0.1\). Thus, during this early stage, the total optical depth is dominated by the C/D$_2$ background. As mixing and shock heating proceed, the W-doped layer is heated into the \(T_e\simeq0.6\)--\(0.8\,\mathrm{keV}\) range between \(t=\SI{2.15}{ns}\) and \SI{2.25}{ns}. At \(E_{photon}=\SI{8460}{eV}\), the tungsten optical depth reaches its maximum within this temperature range. Consequently, \(\tau_\mathrm{W}\) increases from \(\sim0.1\) at \SI{2.05}{ns} to become comparable to \(\tau_{\mathrm{C+D_2}}\) by \SI{2.20}{ns}, and exceeds the C+D$_2$ contribution by nearly a factor of two by \SI{2.25}{ns}. In contrast, at \(E_{photon}=\SI{8700}{eV}\), the W opacity peaks near \(T_e\sim\SI{1}{keV}\), above the temperatures reached in the present implosion conditions. The tungsten optical depth therefore remains smaller at this photon energy, resulting in reduced W-to-background contrast, as evident from the comparison of Figs.~\ref{all-3}(k)--(o) and \ref{all-3}(p)--(t). Thus, combining resonant X-ray absorption with Talbot imaging provides spectral sensitivity through the opacity and spatially resolved measurements of the density and temperature-dependent ionization state of W dopants, offering a means to determine whether the W-doped ablator has mixed into the fuel material during the burn phase.

\section{Summary and Outlook}

In summary, this work introduces resonant X-ray absorption imaging as a diagnostic for probing the mixing and heating of doped-ablator material which mixes with the fuel during the compression phase of inertial fusion implosions, in the event that a hard x-ray XFEL could be in the future co-located at an inertial fusion facility. Numerical simulations demonstrate the feasibility of the approach for Cu-doped CH ablators for direct-drive, and W-doped HDC ablators for indirect-drive, such as used in the recent experimental verification on the NIF of fusion ignition and burn.  Experimentally, the approach is demonstrated by 
measuring the hot-spot heating in a tamped laser-driven Cu wire target at solid density, using Talbot deflectometry to image the resonant absorption of the N-like $\mathrm{Cu}^{22+}$ ion charge state. The total material density is obtained directly from the reconstructed phase map, while the opacity is determined from the corresponding transmission map. Combining these measurements with tabulated atomic data of ionization balance and opacity as functions of temperature enables inference of the density and temperature of hot spot. The approach can be generalized to other dopants, and the applicable range of plasma temperature sensitivity depends on the transition energy of the L- or M-shel shells.  For the $\sim$keV range which is relevant for mix and heating during the compression phase, the L-shell ionization of Cu (observing K-L transitions), and M-shell ionization in W (L-M transitions) are relevant. At the $\sim$10 keV temperatures relevant to fusion burn, L-shell ionization of W, accessible through either K--L or L--M transitions, becomes relevant.

The high spectral brightness of an XFEL, far exceeding the X-ray self-emission from a compressed or burning ICF core~\cite{Jarrott2018,Regan2012,Bachmann2020}, provides high signal-to-background conditions for resonant absorption imaging at photon energies around 10~keV when combined with appropriate X-ray optics. For example, X-ray self-emission in individual dopant lines from a 10 MJ implosion is estimated to be at most $\sim$10~J. Assuming a $10^{-6}$~sr collection optic, such as a 1-mm-diameter CRL positioned 1~m from the source, and a detector at 20~m, chromatic aberration permits micron-scale focusing over a bandwidth of a few tens of eV, approximately matching the $\sim$20 eV SASE bandwidth. Photons with energies beyond this bandwidth can be further suppressed using a Bragg crystal or multilayer mirror placed upstream of the detector. For an XFEL pulse energy of 2~mJ and an attenuation factor of $10$--$100$ through the compressed target, only $\sim \upmu\mathrm{J}$ of self-emission background is anticipated to be superimposed on the 20--200~$\upmu\mathrm{J}$ of resonant absorption image.

These results indicate that resonant absorption imaging using XFEL probing at an implosion facility, could be sufficient to resolve ablator--fuel mix, and provide snapshot measurements of the temperature evolution during fusion burn with micron-scale spatial resolution. This can be achieved with dopant levels commensurate with present ICF shell designs. In addition to resonant X-ray absorption imaging, atomic-scale-sensitive X-ray diffraction (XRD) and nanometer-scale-sensitive small-angle X-ray scattering (SAXS) are demonstrated to probe the formation and mixing kinetics of shock-compressed planar polyethylene terephthalate~\cite{He2022}. By simultaneously employing multiple X-ray diagnostics, including SAXS, XPCI, and RXES, we recently demonstrated full-scale spatiotemporally resolved measurements of solid-density plasma dynamics, capturing preplasma generation, ultrafast heating and ionization, laser-driven blast-wave propagation, and transient surface-return-current-induced compression from sub-picosecond to hundreds-of-picoseconds timescales~\cite{Huang2026}. Extended to spherical ICF capsules, these complementary diagnostics could provide multiscale characterization of high-density-carbon ablator layers, ranging from atomic-scale structure to the nano- and microscale evolution of mix seeded by capsule-fabrication defects and foam remants. Thus, integrating complementary X-ray diagnostics—including imaging, scattering, diffraction, and spectroscopy—into a future laser-driven spherical compression facility coupled to a high-brightness XFEL could provide unprecedented insight into laser-fusion plasmas, accelerate the development of fusion energy, and open new frontiers in high-energy-density science.

\section*{Acknowledgement}
We acknowledge European XFEL in Schenefeld, Germany, for provision of X-ray free-electron laser beamtime at HED SASE2 under proposal number 8042. The authors thank the HiBEF user consortium and their staff for their user supports and the equipment provided to make this experiment possible. We acknowledge Tilo Döppner, Benjamin Bachmann, Gareth Hall and Tao Tao for useful discussions. The work of G. P.-C. has been supported by Research Grant No. PID2022-13 7632OB-I00 from the Spanish Ministry of Science and Innovation.

\section*{Authors contributions}
L.H. led the experiment. L.H., L.Y., A.L.G., O.S.H., M.M., C.B., E.B., Y.C., S.D.C., S.G., M.-L.H., H.H., T.K., J.L., M.Masruri,  M.N., M.O., \"{O}.\"{O}., P.O., X.P., F.P.-B., A.P., T.R.P., C.Q., S.V.R, L.R., M.R., H.-P.S., S.S., J.-P.S., G.S., M.S., M.T.,  T.T., and T.E.C. performed the experiment. L.H., L.Y., A.L.G., O.S.H., Y.C., W.F., T.K., A.L.G., J.L., P.O., X.P., G.P.-C, T.R.P., L.R., S.S., M.S., T.T., U.Z., and T.E.C. analyzed the experimental data. L.H. performed hydrodynamic simulations for the direct-drive case, while L.Y. performed hydrodynamic simulations for the indirect-drive case. L.Y. and L.H. wrote Sec.~III.B, “Imaging W-doped shell in the indirect-drive case,” while L.H. wrote the remainder of the original manuscript draft. All authors discussed the results and contributed to revising the manuscript.

\section*{Data availability}

The data recorded for the experiment at the European XFEL are available at EuXFEL data repository, HED 8042, after the expiration of the embargo period~\cite{Data8042} or from the corresponding author L.H. upon request. 

\section*{Code availability}

The MHD code FLASH used in this work was developed in part by the DOE NNSA- and DOE Office of Science–supported Flash Center for Computational Science at the University of Chicago and the University of Rochester, and it is publicly available.

\subsection*{Competing interests}
The authors declare no competing interests.

\bibliographystyle{apsrev4-2}
\bibliography{main}

\end{document}